\documentclass[aps,prb,groupedaddress,preprint,showpacs]{revtex4}
\usepackage{amsfonts}
\usepackage{graphicx}
\usepackage[usenames]{color}

\begin{document}

\title{Wire network behavior in superconducting Nb films with diluted triangular  arrays of holes}


\author{S. K. He}
\author{W. J. Zhang}
\author{H. F. Liu}
\author{G. M. Xue}
\author{B. H. Li}
\author{H. Xiao}
\author{Z. C. Wen}
\author{X. F. Han}
\author{S. P. Zhao}
\author{C. Z. Gu}
\author{X. G. Qiu}

\email[]{xgqiu@aphy.iphy.ac.cn}
\affiliation{Beijing National Laboratory for Condensed Matter Physics, Institute of Physics, Chinese Academy of Science, Beijing 100190, China}


\date{\today}

\begin{abstract}
  We present transport measurement results on superconducting Nb films with diluted triangular arrays (honeycomb and kagom\'{e}) of holes. The patterned films have large disk-shaped interstitial regions even when the edge-to-edge separations between nearest neighboring holes are comparable to the coherence length. Changes in the field interval of two consecutive minima in the field dependent resistance $R(H)$ curves are observed.  In the low field region, fine structures in the $R(H)$  and  $T_c(H)$ curves are identified  in both arrays. Comparison of experimental data with calculation results shows that these structures observed in honeycomb and kagom\'{e} hole arrays resemble those in wire networks with triangular and $T_3$ symmetries, respectively.  Our findings suggest that even in these specified periodic hole arrays with very large interstitial regions, the low field fine structures are determined by the connectivity of the arrays.
\end{abstract}


\maketitle

\section{INTRODUCTION}
Superconducting films with periodic arrays of artificial pinning sites have been extensively studied for a long time.\cite{baert_prl_95,Martin_prl_1999,Hoffmann_prb_2000,Karapetrov_prl_2005} It is found that at the fields where the number of superconducting flux quantum $\Phi_0=hc/2e$ in unit area is an integer multiple of the pinning sites, the so called commensurate effects such as peaks
in the $I_c(H)$ and dips in the $R(H)$ curves can be observed.\cite{Hoffmann_prb_2000,VanLook_prb_2002,Martin_prl_1997} The most prevailing explanation for these effects is that the vortex lattice is commensurate with the underlying array and pinned efficiently at the matching fields.

However, similar phenomena  have also been observed in other systems including superconducting wire networks and Josephson junction arrays.\cite{Pannetier.wire,ling.wirenetwork,jja.Tinkham} In wire networks, when the magnetic flux through single plaquette does not equal to integer multiple of $\Phi_0$, supercurrents must be induced along the loops to satisfy the fluxoid quantization condition. There is an energy cost due to the induced supercurrents and thus results in a reduced $T_c$. From this point of view, in wire networks, the resistance oscillations are caused by $T_c$ suppression  at non-matching fields rather than enhanced pinning at the matching fields.\cite{patel.prb} Interestingly, transitions from the pinning regime to the wire network regime can be observed in some superconducting film with pinning arrays. For example, with increasing hole diameter, the width of the strips between neighboring holes in a hole array becomes comparable to the coherence length at temperatures close to $T_{c0}$ (zero resistance transition temperature) and the system behaves like a wire network.\cite{Bruynseraede,Moshchalkov.wire,Hoffmann_prb_2000}
\begin{figure}
  \includegraphics[width=8cm]{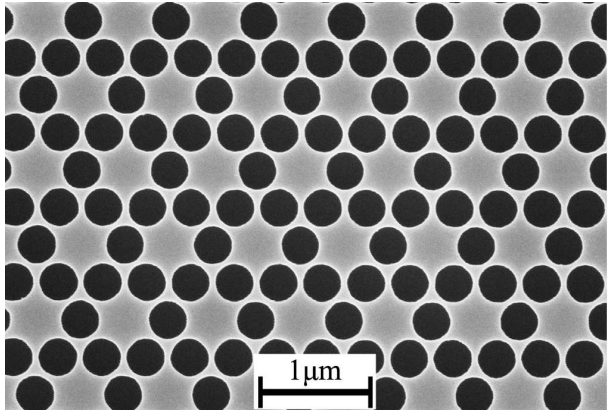}\\
  \caption{Scanning electron microscopy (SEM) image of the superconducting Nb film with a kagom\'{e}  array of holes. The center-to-center distance between nearest neighbor is  400~nm and the hole diameter is about 340~nm.}\label{semkagome}
\end{figure}
We notice that these findings are based on square\cite{Bruynseraede,Moshchalkov.wire,Hoffmann_prb_2000} and triangular\cite{patel.prb} lattices. In those cases, with small edge-to-edge separation, the remaining geometrical structures of the patterned films, are indeed like wire networks with uniform width and small nodes. Thus the transition from pining array to wire network is straightforward. However, other series of hole arrays with large hole diameter may not have such direct geometrical simplification. Investigations of those hole arrays are needed to better understand the physics related to superconductors with micro/nano structures.

In this work, superconducting Nb films with honeycomb and kagom\'{e} arrays of holes are studied.  The arrays can be viewed as diluted triangular arrays, for they can be constructed with 1/3 and 1/4 of the sites removed form the original lattice, respectively.\cite{Reichhardt_prb_2007} The edge-to-edge separations between neighboring holes in these samples are comparable to the coherence length at temperature close to $T_{c0}$. Even though, unlike square and triangular arrays,  the patterned films still have large interstitial regions.
This distinct character makes the system  different from wire networks, for in ideal wire networks, both the width of the stripes and the radius of the nodes are smaller than the coherence length and a uniform order parameter in the cross section of any stripe is expected. Large interstitial regions, on the other hand, can facilitate the nucleation of Abrikosov vortices\cite{Martin_prl_1999,Harada_sci_96,Karapetrov_prl_2005,Metlushko_prb_1999,Wu_jap_2005,Kramer_prl_2009,Grigorenko_prl_2003} , resulting in the appearance of normal cores.
 In our experiments, a series of minima are observed in the $R(H)$ curves. While the oscillation period at low fields is in good agreement with the value derived from the hole density, the periodicity in higher fields is much larger.  It is found that the transition of the two regions is due to the presence of interstitial vortices in the high field region. Surprisingly, in the low field region, wire network behaviors are observed in both samples.

We identify the wire network behavior by the fractional matchings (fine structures) in the $R(H)$ and $T_c(H)$ curves. The positions and the relative values of the matching minima are studied in detail.  We notice that the connectivity of a wire network determines the characters of the fine structures, therefore we highlight the connectivity of our hole arrays and simplify each of them to a wire network.  The comparison of the results in the holes arrays with those of their corresponding wire networks, including reported experimental data as well as calculation results based on the Alexander model,\cite{Alexander.equation} demonstrates that the hole arrays studied in this work are well described by wire networks when subjected to small field.
\section{EXPERIMENT}
The nano-structured superconducting films were prepared as follows.
First, the superconducting Nb film with a thickness of 100~nm was
deposited by magnetron sputtering on Si substrate with SiO$_{2}$ buffer layer. Next, a micro-bridge for four terminal transport measurement was fabricated by ultraviolet photolithography followed by reactive ion etching. Then the desired  arrays  covering the whole bridge
area of $60\times 60~\mu$m$^2$ was patterned by electron-beam
lithography on a polymethyl metacrylate (PMMA) resist layer.
Finally, the pattern was transferred to the Nb film by magnetically enhanced
reactive ion etching. In both the honeycomb and kagom\'{e} samples, the value of the center-to-center distance between nearest neighbor ($a$) is 400~nm and the hole diameter ($d$) is about 340~nm. The scanning electron micrograph image of the kagom\'{e} sample is shown in Fig.~\ref{semkagome}. The smallest width of the stripes between the adjacent holes is about 60~nm.

The transport measurements were carried out in a commercial Physical Properties Measurement System (PPMS) manufactured by Quantum Design. The magnetic field was applied perpendicular to the film surface. During the measurements, the temperature stability was better than 2~mK. The zero field transition temperatures $T_c(0)$ are 8.713~K for the honeycomb sample and 8.755~K for the kagom\'{e} sample, using a criterion of half the normal state resistance $R_N$ at 9~K, which are 8.18~$\Omega$ and 9.02~$\Omega$, respectively. The transition width of these two samples is about 0.15~K. The reference film without any pattern has a slightly higher $T_{c}$ of 8.87~K and transition width of 50~mK. We have measured the $T_{c}(H)$ phase boundary of the reference sample and obtained the zero-temperature coherence length $\xi (0)=9.9$~nm.\cite{Tinkham_book}
\section{RESULTS AND DISCUSSIONS}
\subsection{Reconfiguration}
\begin{figure}
  \includegraphics[width=8cm]{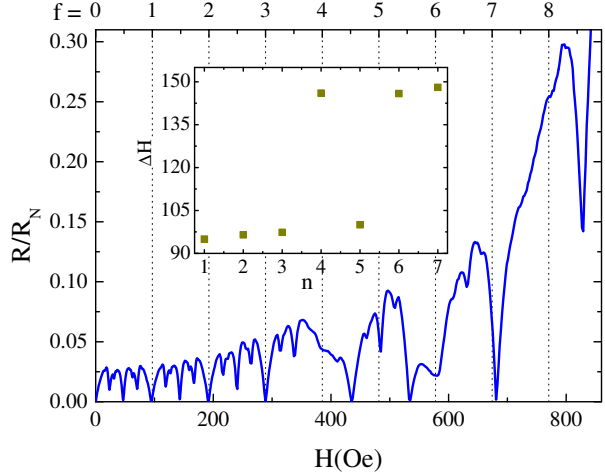}\\
  \caption{(Color online) $R(H)$ curve of  Nb film with honeycomb array of holes measured at 8.55~K with a current of 50~$\mu$A. f is the filling ratio indicating the flux quanta per unit cell. The inset shows the field spacing of two consecutive minima as a function of the minima index.}\label{Rhhoneycomb}
\end{figure}

Figure~\ref{Rhhoneycomb} shows the  $R(H)$ curve measured at 8.55~K with a current of 50~$\mu$A. A series of local minima and maxima are observed. Among these, seven integer matching minima can be clearly identified.  The field interval between two consecutive integer minima versus the corresponding index number are shown in the inset. At fields below 300~Oe, the observed intervals are about 96.5~Oe, is in agreement with the value 99.6~Oe derived from the hole density. When the field is larger than the third matching field, the field spacing has a larger value of about 146.5~Oe. This value is very close to the matching field of a triangular lattice with
the same lattice constant which is 149.4~Oe.  One exception is that the spacing between the 4th and 5th minima is 100~Oe.

\begin{figure}
  \includegraphics[width=8cm]{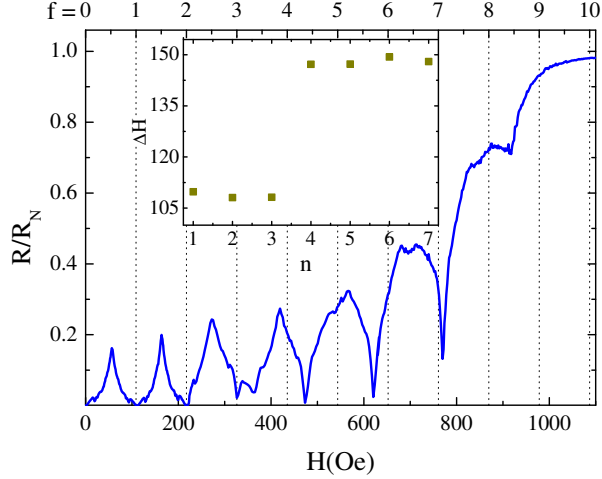}\\
  \caption{(Color online) $R(H)$ curve of  Nb film with kagom\'{e} array of holes measured at 8.60~K with a current of 10~$\mu$A. f is the filling ratio indicating the flux quanta per unit cell. The inset shows the field spacing of two consecutive minima as a function of the minima index.}\label{Rhkagome}
\end{figure}

Figure~\ref{Rhkagome} is the $R(H)$ curve of the kagom\'{e} sample measured at 8.60~K with a current of 10~$\mu$A. Several resistance minima are identified. As can be seen more clearly from the inset of Fig.~\ref{Rhkagome}, the field intervals have two sets of values too. The field intervals of the first three minima are about 108.7~Oe, again in good agreement with the first matching field ($H_1=$112.1~Oe) of the underlying kagom\'{e} lattice. However, the value of the intervals jumps to about 148~Oe when the field is larger than the third matching field.

The change in the field interval of two consecutive minima
have been observed in rectangular arrays and been explained
in terms of reconfiguration of the vortex lattice
from rectangular to square.\cite{Martin_prl_1999,Martin_prb_2000} The interstitial regions in rectangular arrays are stripes which lead to the channeling effect of vortices.\cite{PhysRevB.65.104511,PhysRevB.68.224504}
While in honeycomb and kagom\'{e} arrays the interstitial regions have disk-like shape.  At the typical measurement temperatures 8.60~K,
the diameters of the interstitial regions, approximately 520~nm, are about six times larger than the coherence length.  Thus, Abrikosov vortices can be easily accommodated  by the interstitial sites.

Form the field spacing values between two consecutive integer matchings, the overall feature of the $R(H)$ curves can be explained as follows. When the field is small, supercurrents are generated around the holes in order to satisfy the fluxoid quantization condition. At the integer matching field, it is the arrangement of the fluxoids that is commensurate with the hole lattice.
When the applied magnetic field is larger than the third matching field, Abrikosov vortices are generated in the interstitial region to minimize the free energy.\cite{Wu_jap_2005} These interstitial vortices are effectively pinned by the repulsive interactions from the fluxoids in the holes, which create a caging potential.\cite{Berdiyorov_prl_2006,PhysRevB.79.134501} The overall flux lattice, which is composed of interstitial vortices and the fluxoids in the holes, would have a triangular symmetry. That is why the field intervals of the minima observed in high field region in both lattices are in good agreement with the value derived form a vortex lattice with triangular arrangement. The saturation number $n_s$, namely, the maximum number of fluxoids trapped by one hole without the entry of interstitial vortices are three in both samples. This value exceeds the limit $n_s=d/(4\xi(T))\simeq 1$ given by Mkrtchyan and Schmidt,\cite{Mkrtchyan.jetp} where $d$ is the diameter of the defect and $\xi(T)$ is the temperature dependent coherence length. This fact indicates that the proximity of the holes is prominent.\cite{Ns.doria}

However, the exact vortex configurations  and their evolution are strongly dependent on the density of flux quanta and the  confinement  of the mesoscopic structure. Both image experiment
\cite{Karapetrov_prl_2005} and simulation \cite{Reichhardt_prb_2006,Berdiyorov_prl_2006,Berdiyorov_prb_2006}
have revealed the complex nature and the transition process between various configurations of the composite vortex lattice. This may also responsible for the observation of a field interval of 100~Oe between the 4th and 5th minima in Fig.~\ref{Rhhoneycomb}.
\subsection{Fractional matching and wire network behavior}
In the low field region, fine and repeatable sub-minima are observed as can be seen from Fig.~\ref{Rhhoneycomb} and Fig.~\ref{Rhkagome}.  Because no interstitial vortex is involved in this region and the proximity of the holes, we can related the systems to wire networks despite the large disk-shaped interstitial sites.

\begin{figure}
  \includegraphics[width=8cm]{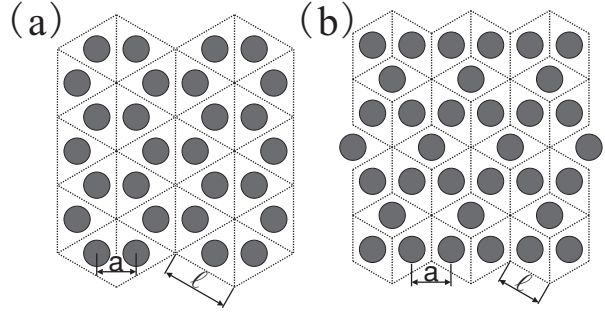}\\
  \caption{Illustration of the transformation from hole arrays (dark circles) to wire networks (dotted lines).  (a) Honeycomb hole array to triangular wire network. (b) Kagom\'{e} hole array to $T_3$  wire network. $a$: the distance between the centers of the neighboring holes. $l$: the side length of the corresponding wire networks.}\label{transform}
\end{figure}

 We notice that a  wire network is obtained by assigning nodes to the center of the interstitial regions in the original hole array and connecting them. In Fig.~\ref{transform}, the centers of the interstitial regions are regarded as nodes. Then, the superconducting stripes between neighboring holes are viewed as wires (dotted lines) connecting the nodes.  Following this procedure, the connectivity of the patterned films are highlighted and the honeycomb and kagom\'{e} hole arrays are transformed to triangular and $T_3$ wire networks, respectively. Although C.~C.~Abilio \textit{et al}. have found that superconducting Al film with a square array of holes can be described as a square wire network,\cite{Abilio.JLow} the simplification we made here is more radical for the diameter of the disk-shaped interstitial regions is even larger than that of the holes.
Assuming that the distance between the centers of the neighboring holes is $a$ in the hole array, then in the dual lattices the side length $l$ of the elementary triangles will be $\sqrt3 a$  and the side length of the rhombus tile in the $T_3$ geometry will be $2\sqrt 3 a/3$ as can been seen from Fig.~\ref{transform}(a) and Fig.~\ref{transform}(b), respectively. If the matching effects of the hole arrays originate from $T_c$ suppression at non-matching fields, the experimental results should show similar features with  what have been observed in their corresponding wire networks.

\begin{figure}
  \includegraphics[width=8cm]{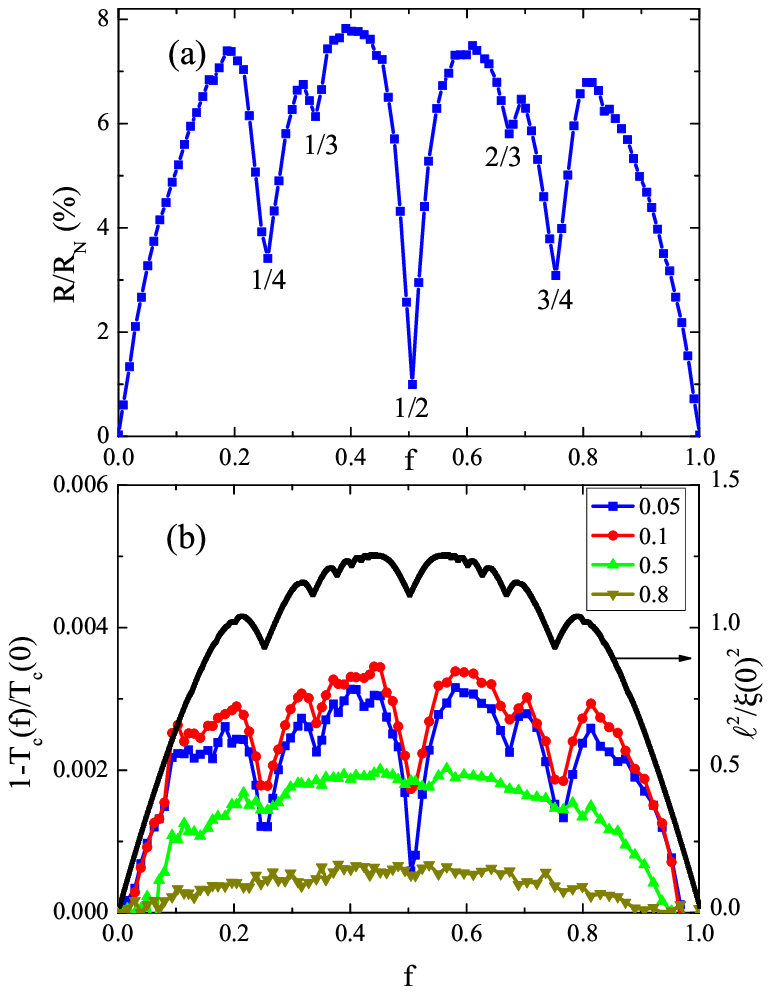}\\
  \caption{(Color online) (a)Low field  $R(H)$ curve for the sample with a honeycomb hole array measured at 8.60~K and $I=$10~$\mu$A. The field is normalized by the first matching field and the resistance is normalized by $R_N$ at 9~K.  (b)The field dependent transition temperature $T_c(H)$ curves of the honeycomb sample with different criteria $r=R/R_N$. The parabolic background has been subtracted. The top one is the theoretical curve for triangular wire network in unit of $l^2/\xi(0)^2$. }\label{honeycomb}
\end{figure}
Figure \ref{honeycomb}(a) shows the fine structures of the $R(H)$ curve of the honeycomb sample measured at 8.60~K with a current of 10~$\mu$A. The $x$ axis is given as filling ratio, $f=\Phi/\Phi_0$, where $\Phi=Ha^23\sqrt 3/4$ is the magnetic flux per elementary triangle of the corresponding wire network. In the field range $0\leq f \leq 1$,  resistance minima are observed at $f=$ 1/4, 1/3, 1/2, 2/3 and 3/4.  The magnitude of the oscillation is about eight percent of $R_N$. At the field $f=$1/2 where the commensurate effect is most pronounced, the resistance drops to a value comparable to that at zero field. At $f=$ 1/4 and 3/4, the resistances are about three percent of $R_N$.

Figure \ref{honeycomb}(b) shows the $T_c(H)$ curves in reduced units, $\Delta T_c/T_{c}(0)$  vs $f$,  using different resistance criteria $r=R/R_N$.  The curves were obtained with a current of 10~$\mu$A. The parabolic background\cite{Pannetier.wire} which reflects the contribution of the finite width of the strands to the critical field has been subtracted.
For $r=$0.05 and 0.1, dips  are observed at filling ratios $f=$ 1/4, 1/3, 1/2, 2/3 and 3/4. With the  criterion of $r=$ 0.5, only dips at $f=$ 1/4, 1/2 and 3/4 are visible and those at $f=$ 1/3 and 2/3 are missing. For $r=$ 0.8, $T_c$  changes smoothly with field and no features can be identified. Thus, comparing Fig.~\ref{honeycomb}(a) and (b), one can see that minima in the $R(H)$ curve is located at the same fields where dips are observed in the $T_c(H)$ curve. The fine structures are more pronounced in $R(H)$ than in $\Delta T_c/T_{c}(0)$  vs $f$. This is because the resistance is measured at temperatures slightly above $T_{c0}$ in the superconducting transition region where resistance drops sharply with decreasing temperature, a small change in $T_c$ will result in an enhanced effect in resistance, and thus a more pronounced commensurate effect.

To our knowledge, fractional matching effects in honeycomb arrays have only been observed at $f=$ 1/2 in previous studies.\cite{Wu_jap_2005,Reichhardt_prb_2007} At the typical measuring temperature $T=$ 8.6K (0.987 $T_c$) in this work, $\xi(T)=\xi(0)/\sqrt {1 - T/{T_c}}=$ 86.9~nm is slightly larger than the width of the narrow stripes. This fact makes the samples investigated here be qualitatively different from those in previous works on hole arrays and be better described as wire networks. Rich structures in the $T_c(H)$ curves of wire networks may serve as fingerprints  to differentiate one array from another, since for a given geometry the fine structures are only expected at a particular series of filling ratios.\cite{Pannetier.wire,nori.wire,PhysRevB.82.134532}
From the linearized Ginzburg-Landau equations, Alexander had derived the relation of the order parameters at the nodes subjected to non-integer flux.\cite{Alexander.equation} Based on those equations, the task of finding the field dependent transition temperature is reduced to eigenvalue problems. The mathematical treatment is very similar to that of the tight bounding electrons in 2D arrays subjected to an external field which leads to the famous Hofstadter butterfly energy spectrum.\cite{Hofstadter} The top curve plotted in Fig.~\ref{honeycomb}(b) is the theoretical values of $T_c(H)$ of the triangular wire network, which is the dual structure of honeycomb hole array. The curve is  calculated by
\begin{equation}
\frac{{{T_c}(0) - {T_c}(f)}}{{{T_c}(0)}} = \frac{{\xi {{(0)}^2}}}{{{l^2}}}{\left( {\arccos \frac{{{\varepsilon _t}}}{6}} \right)^2},
\end{equation}
where $l$ is the side length of the wires, $\varepsilon _t$ is the eigenvalue\cite{Alexander.equation} of the following equation obtained by using Landau gauge and  periodic boundary conditions,
\begin{eqnarray}
\varepsilon_t {\psi _n} &=& 2\cos \left[ {\pi (2n - 1)f - {k_y}/2} \right]{e^{ - i{k_y}/2}}{\psi _{n - 1}}\nonumber\\&+&2\cos \left[ {4\pi nf - {k_y}} \right]{\psi _n} \nonumber\\&+&2\cos \left[ {\pi (2n + 1)f - {k_y}/2} \right]{e^{i{k_y}/2}}{\psi _{n + 1}},
\end{eqnarray}
where $k_y=2\pi \frac{{k - 1}}{N},\quad  k = 1,...,N$ implies the periodic condition, $n$ denotes the node index and $\psi _n$ is the order parameter at node $n$.
The fine structures of the $T_c(H)$ of triangular wire network have also been studied by analytical approach based on multiple-loop Aharonov-Bohm Feynman path integrals.\cite{nori.wire} The main features in $T_c(H)$ such as the position and relative strength of the most prominent dips are the same as those in the Alexander's treatment. The most pronounced dips occur at $f=$ 1/4, 1/3, 1/2 etc., in good  agreement with the experimental data. At the fields where dips are observed, the order parameter at the nodes interferences constructively and form different locked-in states corresponding to local maxima in $T_c$.\cite{Higgins.kagome}  In the fabrication process, contamination and damage of the samples are most significant to the narrow strips, resulting in a lower $T_c$ and a widening in the transition width compared to the reference film. The phase coherence between adjacent nodes is fully established when all the stripes are in superconducting state. Therefore, the interference effects is strong near the zero resistance state $r=$ 0.05  and significantly reduced for $r>0.1$.

\begin{figure}
  \includegraphics[width=8.5cm]{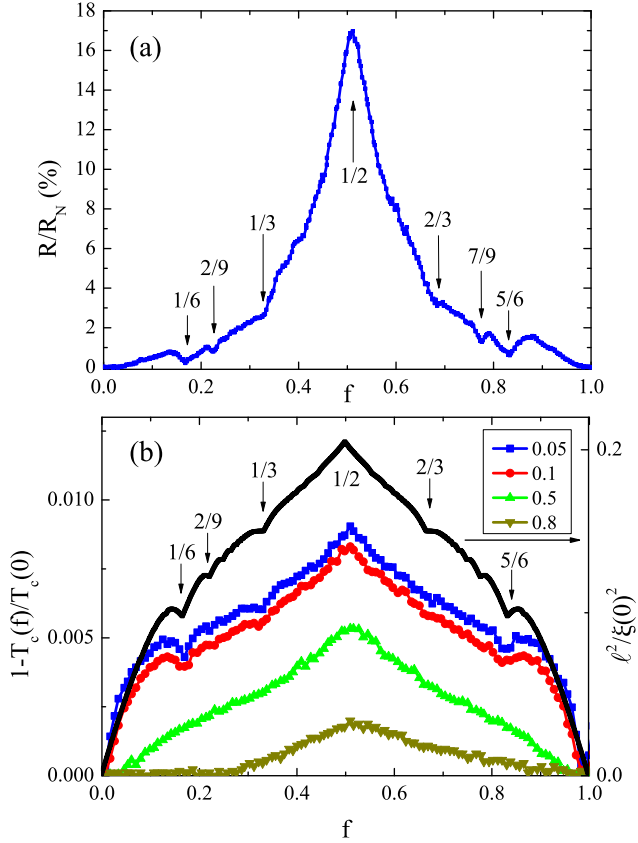}\\
  \caption{(Color online) (a)Low field  $R(H)$ curve for the sample with a kagom\'{e} array measured at 8.67~K and $I$=10~$\mu$A. (b) The field dependent transition temperature $T_c(H)$ of the kagom\'{e} sample with different criteria $r=R/R_N$. The parabolic background has been subtracted. The top one is the theoretical curve for $T_3$ wire network (Ref.~\onlinecite{abilio.t3}) in unit of $l^2/\xi(0)^2$.}\label{kagome}
\end{figure}

Figure \ref{kagome}(a) is the low field $R(H)$ curve of the kagom\'{e} sample measured at 8.67~K. With kagom\'{e} geometry, $f$ corresponds to the magnetic flux through a rhombic tile with side length $2\sqrt 3 a/3$:  $f = {\Phi }/{{\Phi _0}} = {2\sqrt 3 {a^2}H}/{(3\Phi _0)}$.  Again we focus on the results for $f$ in the interval between 0 and 1. Dips are identified  at $f$=1/6, 2/9, 7/9 and 5/6. These fine structures  are observed for the first time in kagom\'{e} hole array, perhaps because of the specified geometrical parameters. Anomalies are also visible at $f=$ 1/3 and 2/3, but they  are more like kinks than dips. Most strikingly, in contrast to the dip observed in the honeycomb sample, a remarkable peak is observed at $f=$ 1/2. The resistance is far above the background, reaching 16.9 percent of the normal state resistance.

Figure \ref{kagome}(b) shows the $T_c(H)$ curves in reduced units, $\Delta T_c/T_{c}(0)$  vs $f$, of the kagom\'{e} sample determined by using different criteria of $r$. Again, all the measurement were carried out with a current of 10~$\mu$A. For small value of $r$ (0.05 and 0.1), dips present at 1/6, 2/9, 1/3, 2/3 and 5/6, although the one at 2/3 is relatively weak (see the labels in Fig.~\ref{kagome}(b)). For larger values of $r$ (0.5 and 0.8), the dips at 1/6, 2/9 and 1/3  gradually disappear but the peak at $f=$ 1/2 remains pronounced.


Most of the fractional matchings observed here are absent in previous works performed on kagom\'{e} lattice.\cite{morgan_prl_98,PhysRevB.64.104505,Reichhardt_prb_2007,PhysRevB.79.134501,kagome1over3} In a recent work on hole arrays with small hole size which is in the pinning regime limit, clear fractional matchings were only observed at $f=$ 1/3 and 2/3.\cite{kagome1over3} Contrasting to their findings, in our work with edge-to-edge separation comparable to the coherence length, dips at $f=$ 1/6 and 5/6 are most pronounced and a distinct peak anomaly at $f=$ 1/2 is observed. The results agree well with what have been observed in $T_3$ wire networks,\cite{abilio.t3,Vidal.kagome} the dual structure of the kagom\'{e} hole array as seen in Fig.~\ref{transform}(b). The top curve of Fig.~\ref{kagome}(b) shows the theoretical curve of $T_3$ networks.\cite{abilio.t3} This is done by the following equation which relates the eigenvalues of $T_3$ ($\varepsilon(f)$) at filling ratio $f$ to the eigenvalues for triangular lattice at $3f/2$.\cite{abilio.t3}
\begin{equation}
{\varepsilon ^2}(f) - 6 = 2\cos (\pi f){\varepsilon _t}(3f/2)
\end{equation}
The comparison with the theoretical curve shows that only the dips at $f$=7/9 and other weaker ones are absent in the experimental curves. For $f$=1/6, 2/9 and 1/3, constructive interference occurs among the superconducting order parameters of the interstitial sites and the fluxoids establish long range locked-in  commensurate order. While for $f$=1/2, superconductivity is localized in single tiles and long range coherence between network sites cannot be established.\cite{abilio.t3}  This kind of fully destructive quantum interference has also been observed in  kagom\'{e} wire networks.\cite{Higgins.kagome,chaikin.PhysRevB} The oscillation of $R(H)$ curve with a peak at half a flux per tile is similar to the single-loop Little-Parks effect. In that case, the supercurrent density reaches the largest value at half flux in order to satisfy the fluxoid quantization condition.\cite{little.parks}

The results obtained from these two lattices confirms the validity of the transform from a hole array to a wire network. However, in contrast to the small nodes in ideal wire networks or square arrays of holes, the geometries studied here possess very large superconducting disk-shaped nodes which can trap  Abrikosov vortices easily. The formation of Abrikosov vortices implies spatial variation and presence of zero points of the order parameters in the interstitial region. Then the simplification of the large interstitial region to a node in a wire network is no longer appropriate. Therefore, transformation from hole arrays to wire networks is only valid for small field values.

\section{CONCLUSIONS}
In conclusion, we have studied the commensurate effects in superconducting films  with honeycomb and kagom\'{e} arrays of holes with small edge-to-edge separation. We found that the magnetoresistance curves have two regions with different oscillation periods. At small fields, there are no Abrikosov vortices presents in the interstitial regions. The large disk-shaped interstitial region can be simplified as a single node and a one-to-one correspondence between hole arrays and wire networks is established. Comparison of experimental data with calculation shows that the simplification works well. Our results suggest that at low fields, the behavior of these specified periodic hole arrays are determined by the connectivity of the systems.
\section{ACKNOWLEDGEMENTS}
We thank Q.~Niu, X.~C.~Xie, V.~V.~Moshchalkov, B.~Y.~Zhu and Y.~Yeshurun for fruitful discussion. This work is supported by National Basic Research Program of China(No.2009CB929100,~2011CBA00107,~2012CB921302) and National Science Foundation of China (No.~91121004,~10974241,~11104335).

\end{document}